\documentclass[aps,showpacs,prc,nofootinbib,floatfix]{revtex4}

\usepackage{graphicx}

\usepackage{latexsym}
\usepackage{hhline}
\usepackage{amssymb}
\usepackage{wasysym}
\usepackage{epsfig}
\usepackage{ulem}
\usepackage{bbm}
\usepackage{yfonts}
\sloppypar

\DeclareMathAlphabet{\mathpzc}{OT1}{pzc}{m}{it}

\newcommand{\be}{\begin{equation}}
\newcommand{\ee}{\end{equation}}
\newcommand{\bea}{\begin{eqnarray}}
\newcommand{\eea}{\end{eqnarray}}
\newcommand{\ba}{\begin{array}}
\newcommand{\ea}{\end{array}}
\newcommand{\bi}{\begin{itemize}}
\newcommand{\ei}{\end{itemize}}
\newcommand{\mi}{\mbox i}

\newcommand{\refe}[1]{(\ref{#1})}
\newcommand{\dleft}{\overleftarrow{\slash\partial}}    
\newcommand{\dright}{\overrightarrow{\slash\partial}}

\newcommand{\mbb}[1]{\mathbbm{#1}}    
\newcommand{\drleft}[1]{\overleftarrow{\partial^{#1}}}    
\newcommand{\drright}[1]{\overrightarrow{\partial^{#1}}}

\newcommand{\drup}[1]{\partial^{#1}}    
\newcommand{\drdw}[1]{\partial_{#1}}

\newcommand{\mt}{{\rm g}}

\newcommand{\mcl}{{\mathcal L}}

\renewcommand{\slash}{/ \!\!\!\!\,}

\renewcommand{\ll}{ /\hspace*{-2.1mm}
}

\newcommand{\foh}{\frac{1}{2}}

\newcommand{\fth}{\frac{3}{2}}
\newcommand{\ffh}{\frac{5}{2}}

\begin{document}

\title{Spin-$\ffh$ fields in  hadron physics.
\footnote{Supported by DFG, contract Le 435/7-1 and SFB/TR16, project B7 }}

\author{V. Shklyar\footnote{On leave from Far Eastern State University,
                       690600 Vladivostok, Russia}}
\email{shklyar@theo.physik.uni-giessen.de}
\author{H. Lenske}
\author{U. Mosel}
\affiliation{Institut f\"ur Theoretische Physik, Universit\"at Giessen, D-35392
Giessen, Germany}

\begin{abstract}
We show that the Lagrangian of the free spin-$\ffh$ field in the spinor-tensor representation with
the auxiliary spinor field depends on the three arbitrary parameters.  The first two parameters are associated 
with the spin-$\fth$ and -$\foh$ sector of the theory  while the latter one is related to the auxiliary 
degrees of freedom. We derive a corresponding propagator of the system which represents (2\,x\,2) matrix in the 
$(\psi_{\mu\nu},\xi)$ space. The diagonal terms stand for the propagation of the spin-$\ffh$ and
auxiliary fields whereas the non-diagonal ones correspond to the  $\psi_{\mu\nu}-\xi$ mixing.
The resulting spin-$\ffh$  propagator  contains non-pole contributions coming from the spin-$\fth$
and -$\foh$ sector of the spinor-tensor representation. A general form of the interaction vertex involving 
spin-$\ffh$ field is discussed  on the example of the $\pi N N^*_\ffh$ coupling. It is demonstrated that 
lower spin degrees of freedom can be removed from the theory by using higher order derivative coupling.

\end{abstract}


\maketitle

\section{Introduction}
The description of   pion- and photon-induced reactions in the resonance energy region is faced with  the 
problem of proper treatment of higher spin states. In 1941  Rarita and Schwinger (R-S)  suggested a set of 
equations  which a field function of a higher spin should obey \cite{Rarita:1941}. Another formulation 
has been developed by Fierz and Pauli \cite{Fierz:1939ix}  where an auxiliary field 
concept is used to derive subsidiary constraints on the field function. 

Regardless of the procedure used  the obtained  Lagrangians for free higher-spin fields 
turn out to be always dependent on arbitrary free parameters. For the spin-$\fth$ fields  
this issue is widely discussed in the Literature: 
(see e.g. \cite{Pascalutsa:1998pw,Shklyar:2008kt,Krebs:2008zb} for a modern status  of the problem).
The case of the spin-$\ffh$ fields is less studied.
First  attempts in this way  have been made in 
\cite{Schwinger:1973rv,Berends:1979rv} where a theory of free  fields has been suggested.
The authors of \cite{Berends:1979rv} deduced an equation of motion as a decomposition in terms of 
corresponding projection operators with  additional algebraic constraints on parameters of the decomposition. 
Schwinger \cite{Schwinger:1973rv} derived a particular form of the spin-$\ffh$ equation
which  coincides with the equation suggested in  \cite{Berends:1979rv} for a specific
choice of the parameters. 

The free particle propagator is a central quantity in  most of the calculations in  quantum
field theory. 
In  \cite{Berends:1979rv} the authors deduced a spin-$\ffh$ propagator written in 
 operator form.
In practical calculations, however, one needs an explicit expression of the propagator.
An attempt to construct a propagator  only from the spin-$\ffh$ projection operator 
has been made in \cite{David:1995pi}. 
We demonstrate that such a quantity  is not consistent with the equation of motions
for the spin-$\ffh$ field.  Another pathology is experienced with the propagator \cite{Renard:1972vv} and 
projector \cite{Zetenyi:2002jy} used in calculations of the resonance production amplitudes:
they do not fulfill the condition  $[\gamma_0 G^{\ffh}_{\mu\nu;\rho\sigma}]^{\mathrm{\dag}}
=\gamma_0 G^{\ffh}_{\rho\sigma;\mu\nu}$
and consequently are not  hermitian. 
Therefore it is important to derive the propagator and investigate its properties in detail. 
To our knowledge no such a study has been done so far. 

The aim of the paper is to deduce an explicit 
expression for the  spin-$\ffh$ propagator and study its properties. 
Guided by the properties  of the 
free spin-$\fth$  R-S theory one would expect the equation of motion for the spin-$\ffh$ field has 
two arbitrary free parameters  which define the non-pole  spin-$\fth$ and-$\foh$ contributions
to the full propagator. The coupling of the spin-$\ffh$ field to the (e.g.) pion-nucleon final state  is 
therefore  defined up to two 'off-shell' parameters \cite{shklyar:2004a} which scale the non-pole 
contributions to the physical observables. Hence, one can ask whether such an arbitrariness can be 
removed from the theory.

The possibility to construct consistent higher-spin  massless theories has already been 
pointed out by Weinberg and Witten a while ago \cite{Weinberg:1980kq}. 
Pascalutsa has shown that  by using a gauge invariant coupling for 
higher spin fields it is possible  to remove the extra-degrees of freedom \cite{Pascalutsa:1999zz}
in a particular case of the R-S theory which maintains gauge invariance in the massless limit. 

As we demonstrated  in \cite{Shklyar:2008kt} the demand for the gauge-invariance  
may not be  enough to eliminate the extra degrees of freedom at the interaction vertex.
The problem appears when the theory does not have a massless limit. However, a coupling which removes non-pole 
terms from the spin-$\ffh$ propagator can be easily constructed by using higher order derivatives. 
A corresponding interaction Lagragian has been deduced in \cite{Shklyar:2008kt} for the case of 
spin-$\fth$ fields and can be easily  extended to  higher spins too.

The paper is organized as follows: \ in Section  \ref{Free52} we 
suggest an alternative form  of the free spin-$\ffh$ Lagrangian as compared to  \cite{Berends:1979rv}  
and discuss its properties in details. The presence of auxiliary field 
complicates the derivation  of the propagator. Therefore, in Section \ref{FreeVector} we  first demonstrate how 
the free  propagator can be obtained for a vector field in the presence of an auxiliary one.  The method is 
then applied to the spin-$\ffh$ field, Section \ref{FreeF}. The resulting spin-$\ffh$ propagator contains
contributions corresponding to the lower spin-$\fth$,\,-$\foh$ sector of the spin-tensor representation. 
In  Section \ref{Coupling} we discuss how these degrees of freedom can be removed from the physical observables 
the the example of the pion-nucleon scattering amplitude. The results are summarized in Section   \ref{Summary}.

\section{\label{Free52} Free spin-$\ffh$ field.}
The field function of higher spins in a spinor-tensor  representation  is a solution of  the set of equations
suggested by  Rarita and Schwinger in \cite{Rarita:1941}. 
In a consistent theory the description of the free field is specified by setting up 
an appropriate Lagrange function $\mathcal{L(\psi_{\mu\nu},\partial_{\rho} \psi_{\mu\nu})}$. 
The spin-$\ffh$ Lagrangian in the presence of the auxiliary spinor field $\xi(x)$ can be written 
in the form
\bea
\mathcal{L} =\mathcal{L}^{\mathrm{(\mathrm{1})}}+\mathcal{L}^{\mathrm{(2)}} +\mathcal{L}^{\mathrm{(aux)}},
\label{applagr_aux}
\eea
where the explicit expressions for $\mathcal{L}^{\mathrm{(\mathrm{1})}}$, $\mathcal{L}^{\mathrm{(2)}}$, and 
$\mathcal{L}^{\mathrm{(aux)}}$ read
\bea
\mathcal{L}^{\mathrm{(1)}}&=&\mi\, a\,
       \bar\psi_{\mu\nu}(x)\biggl( 
                (\gamma^\mu \mt^{\nu\sigma}+\gamma^\nu \mt^{\mu\sigma})\,\drright{\rho}      
     +   (\gamma^\mu \mt^{\nu\rho}+\gamma^\nu \mt^{\mu\rho})\,\drright{\sigma} \nonumber\\
 &&  -  (\gamma^\rho \mt^{\nu\sigma}+\gamma^\sigma \mt^{\nu\rho})\,\drleft{\mu}
    - (\gamma^\rho \mt^{\mu\sigma}+\gamma^\sigma \mt^{\mu\rho})\,\drleft{\nu}\,\biggl)\psi_{\rho\sigma}(x)\nonumber\\
&+&\mi\,\frac{{F}_1(a)}{2}
     \bar\psi_{\mu\nu}(x)\gamma^\lambda \left( \dright -  \dleft \right)\gamma^\delta \psi_{\rho\sigma}(x)
       \left(\mt^{\lambda\mu}\mt^{\delta\rho}\mt^{\nu\sigma}
       +\mt^{\lambda\nu}\mt^{\delta\rho}\mt^{\mu\sigma}
       +\mt^{\lambda\mu}\mt^{\delta\sigma}\mt^{\nu\rho}
       +\mt^{\lambda\nu}\mt^{\delta\sigma}\mt^{\mu\rho}\right)\nonumber\\
&+&m\, F_2(a)\,
       \bar\psi_{\mu\nu}(x) \left (\gamma^\mu \gamma^\rho  \mt^{\nu\sigma} 
                                  +\gamma^\nu \gamma^\rho  \mt^{\mu\sigma} 
                                  +\gamma^\mu \gamma^\sigma  \mt^{\nu\rho} 
                                  +\gamma^\nu \gamma^\sigma  \mt^{\mu\rho}\right) 
       \psi_{\rho\sigma}(x)\nonumber\\
&+&\bar \psi_{\mu\nu}(x) \left ( \frac{\mi}{2} \left (\dright -\dleft\right) -m\right)\psi_{\rho\sigma}(x)
\left(\mt^{\mu\rho}\mt^{\nu\sigma}+\mt^{\mu\sigma}\mt^{\nu\rho}\right),\nonumber\\
\nonumber\\
\mathcal{L}^{\mathrm{(2)}}&=&
  \mi\,b\, \bar \psi_{\mu\nu}(x)\left( \mt^{\mu\nu}\left(\gamma^\rho\drright{\sigma}+\gamma^\sigma\drright{\rho}\right) 
    -\left(\gamma^\nu\drleft{\mu}+\gamma^\mu\drleft{\nu} \right)\mt^{\rho\sigma}\right )\psi_{\rho\sigma}(x)
      \nonumber\\
&+&\mi \,\frac{G_1(a,b)}{2} \, \bar \psi_{\mu\nu}(x) \mt^{\mu\nu}\left(\dright-\dleft\right)
  \mt^{\rho\sigma}  \psi_{\rho\sigma}(x)
        + m\,G_2(a,b)\,\bar \psi_{\mu\nu}(x) \mt^{\mu\nu} \mt^{\rho\sigma}  \psi_{\rho\sigma}(x),\nonumber\\
\nonumber\\
\mathcal{L}^{\mathrm{(aux)}}&=& m \,c\left( \bar \psi_{\mu\nu}(x) \mt^{\mu\nu}\xi(x) + \bar \xi(x) \mt^{\rho\sigma}  
   \psi_{\rho\sigma}(x) \right)
    +
      B(a,b,c)\,\bar \xi(x)\left(\frac{\mi}{2}( \dright-\dleft) +3m\right)\xi(x),
\label{applagr_nontr}
\eea
and $F_1(a)$, $F_2(a)$, $G_1(a,b)$, $G_2(a,b)$, and $B(a,b,c)$ are functions of the free real parameters $a$, $b$, and $c$,
see Appendix \ref{app_lagr}.

The Lagrangian eq.\,\refe{applagr_aux} in general depends on only three independent real parameters $a$, $b$, and $c$.
This formulation of the spin-$\ffh$ theory   is simpler than that of suggested in 
\cite{Berends:1979rv}. In fact, the Lagrangian in \cite{Berends:1979rv} is written as a decomposition in terms of 
projection operators with a number of free parameters. These paramteres  are subjected to additional subsiduary 
constraints need to be resolved.

Independent variations of  $\psi_{\mu\nu}$ and $\xi$ fields give two equations of motion which in momentum
space can be written in the following form
\bea
\left(\Lambda^{\mathrm{(1)}}_{\mu\nu;\rho\sigma}(p)+\Lambda^{\mathrm{(2)}}_{\mu\nu;\rho\sigma}(p)\right)
\,\psi^{\rho\sigma}(p)+c\,m\,\mt^{\mu\nu}\xi(p)=0,\label{appspin52_eq1}\\
m\, c\, \mt^{\rho\sigma}  \psi_{\rho\sigma}(p) + B(a,b,c)\left( \slash p + 3m \right)\xi(p)=0,
\label{appspin52_eq2}
\eea 
where the operators $\Lambda^{\mathrm{(1)}}_{\mu\nu;\rho\sigma}(p)$, $\Lambda^{\mathrm{(2)}}_{\mu\nu;\rho\sigma}(p)$ are

\bea
\Lambda_{\mu\nu;\rho\sigma}^{\mathrm{(1)}}(p)&=& 
(\slash p -m )(\mt_{\mu\sigma}\mt_{\nu\rho} + \mt_{\mu\rho}\mt_{\nu\sigma})\nonumber\\
&+&a(\gamma_\mu p_\rho \mt_{\nu\sigma} + \gamma_\nu p_\rho \mt_{\mu\sigma} 	
      + \gamma_\mu p_\sigma \mt_{\nu\rho} +\gamma_\nu p_\sigma \mt_{\mu\rho} \nonumber\\
&+&  \,\,\,\,\,\,\,  \gamma_\rho p_\mu \mt_{\nu\sigma} +\gamma_\sigma p_\mu \mt_{\nu\rho} 
      +\gamma_\rho p_\nu \mt_{\mu \sigma}  +\gamma_\sigma p_\nu \mt_{\mu\rho}) \nonumber\\
&+& F_1(a)\,(\gamma_\mu \slash p\gamma_\rho   \mt_{\nu\sigma}
      +\gamma_\nu \slash p\gamma_\rho   \mt_{\mu\sigma}
      +\gamma_\mu \slash p\gamma_\sigma \mt_{\nu\rho} 
      +\gamma_\nu \slash p\gamma_\sigma \mt_{\mu\rho} )\nonumber\\
&+&m\,F_2(a)\,( \gamma_\mu\gamma_\rho   \mt_{\nu\sigma}
      +  \gamma_\nu\gamma_\rho   \mt_{\mu\sigma}
      +  \gamma_\mu\gamma_\sigma \mt_{\nu\rho  }
      +  \gamma_\nu\gamma_\sigma \mt_{\mu\rho  } )\label{lambda_v1}\\
\nonumber\\
 \Lambda^{\mathrm{(2)}}_{\mu\nu;\rho\sigma}(p)&=&
    b\,(\gamma_\mu p_\nu \mt_{\rho\sigma} + \gamma_\nu p_\mu \mt_{\rho\sigma} 
      +\gamma_\rho p_\sigma \mt_{\mu\nu} + \gamma_\sigma p_\rho \mt_{\mu\nu} )\nonumber\\
       &+&\left( \slash p\, G_1(a,b)+m\,G_2(a,b)\right) \mt_{\mu\nu}\mt_{\sigma\rho}.
\label{lambda_v2}
\eea

The equations of motion (\ref{appspin52_eq1},\,\ref{appspin52_eq2}) 
are written in the most general form and are consistent with those of 
defined in \cite{Schwinger:1973rv,Berends:1979rv}. For example, the equation suggested by Schwinger corresponds 
to the choice of parameters  $a=-1$, $b=1$, and $c=-2$. Note that the functions
$F_1(a)$, $F_2(a)$, $G_1(a,b)$, and $G_2(a,b)$ do not contain  the parameter $c$ which reflects independence of 
the spin-$\ffh$ field on  the axuliary degrees of freedom.
The R-S constraints \cite{Rarita:1941}
follow from eqs.\,(\ref{appspin52_eq1},\ref{appspin52_eq2}) with the additional condition $\xi(p)$=0, 
see  Appendix\,\ref{app_lagr}.

It is interesting to note, that the operator $\Lambda^{\mathrm{(1)}}_{\mu\nu;\rho\sigma}(p)$ would
give an equation of motion  $\Lambda^{\mathrm{(1)}}_{\mu\nu;\rho\sigma}(p)\psi^{\mu\nu}=0$ 
for the spin-$\ffh$ fields  provided  $\mt^{\mu\nu}\psi_{\mu\nu}=0$, where the later property
is assumed a priori. However, the corresponding inverse operator  
$[\Lambda^{\mathrm{(1)}}_{\mu\nu;\rho\sigma}(p)]^{-1}$ has  additional non-physical poles in 
the spin-$\foh$ sector. This indicates that the constraint   $\mt^{\mu\nu}\psi_{\mu\nu}=0$ should also 
follow from the equation of motion and cannot be assumed a priori. The second operator 
$\Lambda^{\mathrm{(2)}}_{\mu\nu;\rho\sigma}(p)$ acts only  in the spin-$\foh$ sector
of the spin-tensor representation. This can be checked by a direct decomposition of the operator 
\refe{lambda_v2} in terms of  projection operators given in Appendix \ref{projectors} . 
The same conclusion can be drawn
from the observation that $\Lambda^{\mathrm{(2)}}_{\mu\nu;\rho\sigma}(p)$ is orthogonal
to all $\mathcal{P}^\ffh_{\rho\sigma;\tau\delta}(p)$, $\mathcal{P}^\fth_{ij;\rho\sigma;\tau\delta}(p)$
projection operators, where $i,j=1,2$. Hence the parameter $b$ is 
related only to the spin-$\foh$ degrees of freedom whereas $a$ scales both spin-$\fth$ and -$\foh$ ones.

In practical calculations one needs to know a free propagator corresponding to the spin-$\ffh$ field. 
The derivation of the propagator  becomes complicated in the presence of the auxiliary field degrees of freedom.  
To demonstrate the procedure  it is useful to consider first an example of the   free vector field $\varphi_\mu$ 
in the presence of auxiliary one. In the next Section we outline a general procedure which can be applied to the 
spin-$\ffh$ case.

\section{\label{FreeVector} Free vector field}
The idea to use auxiliary degrees of freedom to describe systems with higher spins was first 
utilized in the original work of Fierz and Pauli \cite{Fierz:1939ix}. As is well known, however, there 
is no need for such complications in the case of spins $J\leq 2$ and $J\leq \fth$. For higher 
spins the use of axiliary degrees of becomes inevitable \cite{Berends:1979rv}. 
Here we consider the case of the vector 
field $\varphi^\mu$ in the presence of the additional scalar field $\lambda$ and derive the 
free propagator of the system.
The  Lagrangian of the ($\varphi_\mu,\lambda$) system can be written as 
\bea 
\mcl^{\mbb v}= -\frac{1}{2}\left(\drup{\mu}\varphi_\nu\right)\left(\drdw{\mu}\varphi^\nu\right)
+\frac{1}{2}m^2\varphi_\nu \varphi^\nu + a\, m \left( \drdw{\nu} \varphi^\nu\right)\,\lambda  
-\frac{1}{2}a^2\,m^2 \, \lambda^2,
\label{L_v}
\eea
where $\varphi_\mu$ is a vector and $\lambda$ is an auxiliary scalar field and $a$ is an arbitrary 
free parameter. 
Independent variations of
the vector and auxiliary fields produce two equations of motion
\bea
(\Box + m^2)\varphi_\mu  - a\,m \drdw{\mu}\, \lambda =0,\nonumber\\
a\,m \, \drdw{\mu}\varphi^\mu -a^2 m^2 \,\lambda =0.
\label{eq_v1}
\eea
Diagonalization of the system \refe{eq_v1} leads to  the Proca equation for the vector field $\varphi_\mu$
whereas the auxiliary field $\lambda$ vanishes. Although $\lambda =0$ the propagator 
of the system always contains a component associated  with  the auxiliary field and 
$\varphi_\mu-\lambda$ mixing terms.  

To obtain a  propagator for the  
system  of fields ($\varphi_\mu, \lambda$) it is convenient to rewrite eq.\,\refe{eq_v1}  in the matrix form
\bea
\Lambda^{\mbb v}_{\{\mu\nu\}} \Phi^{\mbb{ v}\{\nu\} } =0,
\label{eq_vmatr}
\eea
where 
\bea
\Lambda^{\mbb v}_{\{\mu\nu\}} =
\left(\begin{array}{cc}
(\Box + m^2) \mt_{\mu\nu} & - a\,m \drdw{\mu}\\
a\,m \, \partial_{\nu}& -a^2 m^2 \\
\end{array}\right) \,\,\, {\rm and}\,\,\,
\Phi^{\mbb{v}\{\nu\}}=\left(\begin{array}{c}
\varphi^\nu\\
\lambda \\
\end{array}\right).\label{v_defn}
\eea  
Since the system contains vector and scalar degrees of freedom the Lorentz indices in curly brackets of 
eqs.\,(\ref{eq_vmatr},\ref{v_defn}) are associated with corresponding  tensor and vector elements of 
$\Lambda^{\mbb v}_{\{\mu\nu\}}$ and $\Phi^{\mbb{v}}_{\{\nu\}}$.
 
The inverse operator (propagator) can be obtained as a solution of the following equation
\bea
\Lambda^{\mbb v \,\,\,\,\,\,\mu\}}_{\,\,\{\rho}\,\, G^{\mbb{v}}_{\{\mu\nu\}}
= I^{\mbb v}_{\{\mu\nu\}}\,\delta^4(x-x'),
\label{v_eq_prop}
\eea
where  the propagator $G^{\mbb{v}}_{\{\mu\nu\}}$ and  the unit matrix  $I^{\mbb v}_{\{\mu\nu\}}$ are defined as
\bea
G^{\mbb v}_{\{\mu\nu\}} =
\left(\begin{array}{cc}
G_{\mu\nu}^{(\varphi\varphi)} & G_\mu^{(\varphi\lambda)}\nonumber\\
G_\nu^{(\lambda\varphi)}& G^{(\lambda\lambda)} \\
\end{array}\right) \,\,\, {\rm and}\,\,\,
I^{\mbb v}_{\{\mu\nu\}}=\left(\begin{array}{c}
\mt_{\mu\nu}\nonumber\\
1 \\
\end{array}\right). 
\label{v_defn2}
\eea
The four  components of the matrix $G^{\mbb v}_{\{\mu\nu\}}$  have  simple physical meanings: 
$G_{\mu\nu}^{(\varphi\varphi)}$ and  $G^{(\lambda\lambda)}$ stand for the propagator of
the purely vector and auxiliary scalar  fields, correspondingly, whereas the nondiagonal 
$G_\mu^{(\varphi\lambda)}\nonumber$ and $G_\nu^{(\lambda\varphi)}$ terms are associated with the 
$\varphi-\lambda$ mixing. The solution of eq.\,\refe{v_eq_prop} in the momentum space is
\bea
G^{\mbb v}_{\{\mu\nu\}}(p) =
\left(\begin{array}{cc}
\frac{\left (-\displaystyle \mt_{\mu\nu}+ \frac{p_\mu p_\nu}{m^2}\right) }{\displaystyle p^2-m^2} & 
{\rm i}{\displaystyle\frac{p_\mu}{a\,m^3}}
\label{vvprop}\\
\\
{-\rm i}{\displaystyle\frac{p_\nu}{a\,m^3}}
& {\displaystyle\frac{(p^2-m^2)}{a^2\,m^4}}\\
\end{array}\right).
\eea
The pole $( p^2-m^2)^{-1}$  appears only at the vector component $G_{\mu\nu}^{(\varphi\varphi)}(p)$; this term
 completely coincides with the corresponding expression well known from   quantum  
field theory. The remaining terms in the propagator depend on the free parameter $a$ associated with the 
auxiliary field $\lambda$. From  eq.\,\refe{vvprop} one can conclude  that $G^{(\lambda\lambda)}$ gives contributions 
only off-shell $p^2 \neq m^2$.  Note that the scalar component of the propagating vector field $\varphi_\mu$ mixes 
with the $\lambda$ field which leads to the appearance of the finite non-diagonal components in the 
propagator eq.\,\refe{vvprop}.

Despite the complications related to the introduction of the auxiliary field the description in terms of 
$(\varphi_\mu,\lambda)$ system is completely equivalent to the conventional description 
in terms of the pure vector field. It implies that physical observables do not depend on the free  parameter 
$a$ appearing in the the full propagator \refe{vvprop}. For the free fields this conclusion immediately
follows from the fact that the auxiliary field can be excluded from the   upper equation \refe{eq_v1}. 
It also holds true in the case of interacting fields provided  there is no coupling to auxiliary degrees of freedom.

\section{\label{FreeF} Propagator for the free spin-$\ffh$ field.}
Similar to the procedure described in Section \ref{FreeVector}  it is convenient to rewrite the set 
of equations (\ref{appspin52_eq1},\,\ref{appspin52_eq2}) in  matrix form 
\bea
\left(\begin{array}{cc}
\Lambda^{(\psi\psi)}_{\mu\nu;\rho\sigma}(p) & m\, c\, \mt_{\mu\nu}\\
c\,m\, \mt_{\rho\sigma}  &  \Lambda^{(\xi\xi)}(p)
\end{array}\right) 
\left(\begin{array}{c}
\psi^{\rho\sigma}(p)\\\xi(p)
\end{array}\right)=0,
\label{appspin52_matrix}
\eea
where $\Lambda^{(\psi\psi)}_{\mu\nu;\rho\sigma}(p)=\Lambda^{\mathrm{(1)}}_{\mu\nu;\rho\sigma}(p)
+\Lambda^{\mathrm{(2)}}_{\mu\nu;\rho\sigma}(p)$
and $\Lambda^{(\xi\xi)}(p)= B(a,b,c)\left( \slash p + 3m \right)$. 
While the auxiliary field vanishes on-shell the full propagator should also contain  an off-shell part 
related  with the auxiliary field $\xi(x)$.
Hence, the full  propagator of the system is
\bea
\mathcal{G}^{\{\tau\lambda;\rho\sigma\}}(p)=
\left(\begin{array}{cc}
G_{(\psi\psi)}^{\tau\lambda;\rho\sigma}(p) & G_{(\psi\xi)}^{\tau\lambda;}(p) \\
G_{(\xi\psi)}^{;\rho\sigma}(p) & G_{(\xi\xi)}(p) \\
\end{array}\right) 
\label{appprop_prop}
\eea
and satisfies the equation 
\bea
\left(\begin{array}{cc}
\Lambda^{(\psi\psi)}_{\mu\nu;\tau\lambda}(p) & m\, c\, \mt_{\mu\nu}\\
c\,m\, \mt_{\tau\lambda}  &  \Lambda^{(\xi\xi)}(p)
\end{array}\right) 
\left(\begin{array}{cc}
G_{(\psi\psi)}^{\tau\lambda;\rho\sigma}(p) & G_{(\psi\xi)}^{\tau\lambda;}(p) \\
G_{(\xi\psi)}^{;\rho\sigma}(p) & G_{(\xi\xi)}(p) \\
\end{array}\right) 
=\left(\begin{array}{cc}
I_{\mu\nu}^{\rho\sigma} & 0 \\
0& 1 \\
\end{array}\right),
\label{appprop_matrix}
\eea
where  $I_{\mu\nu}^{\rho\sigma}=\mt_\mu^\rho \mt_\nu^\sigma+\mt_\mu^\sigma \mt_\nu^\rho$.
 The diagonal terms 
$G_{(\psi\psi)}^{\mu\nu;\rho\sigma}(p)$ and $G_{(\xi\xi)}(p)$  are related 
to fields $\psi$ and $\xi$ respectively  whereas the
non-diagonal ones stand for mixing between the auxiliary  spinor field and the 'off-shell'  
spin-$\foh$ component of the spin-$\ffh$ field.  

The propagator of the spin-$\ffh$ field  $G^\ffh_{\mu\nu;\rho\sigma}(p) 
=G^{(\psi\psi)}_{\mu\nu;\rho\sigma}(p)$ 
is  obtained as a solution of the set of equations 
\bea
\Lambda^{(\psi\psi)}_{\mu\nu;\tau\lambda}(p)\, G_{(\psi\psi)}^{\tau\lambda;\rho\sigma}(p) 
   + m\, c\, \mt_{\mu\nu}\,G_{(\xi\psi)}^{;\rho\sigma}(p) &=& I_{\mu\nu}^{\rho\sigma}, \nonumber\\
c\,m\, \mt_{\tau\lambda}  G_{(\psi\xi)}^{\tau\lambda;}(p) + \Lambda^{(\xi\xi)}(p)\,
   G_{(\xi\xi)}(p)&=&1.
\label{propag_eqn}
\eea
In the literature one  sometimes encounters  a propagator defined as
\bea  
G_{\rho\sigma;\tau\delta}'(p)=\frac{\slash p +m}{p^2-m^2+i\epsilon}
\mathcal{P}^\ffh_{\rho\sigma;\tau\delta}(p),
\label{prop_proj52}
\eea
 where 
$\mathcal{P}^\ffh_{\rho\sigma;\tau\delta}(p)$ is a spin-$\ffh$ projection operator 
in the spinor-tensor representation \refe{app_proj}. However, the quatity  defined above does 
not have an inverse and therefore cannot obey  eqs.\,(\ref{propag_eqn}) for any choice of the free 
parameters. This can be shown by replacing the $G_{(\psi\psi)}^{\tau\lambda;\rho\sigma}(p)$ in  
the upper equation (\ref{propag_eqn}) by expression from eq.\,\refe{prop_proj52} and multiplying the 
both sides of the resulting equation from the right by  a projection operator 
$\mathcal{P}^{\fth}_{22;\rho\sigma,\tau\delta}(p)$. Using the general properties of projection 
operators eq.\,\refe{app_property} the obtained expression reduces to
\bea
 m\, c\, \mt_{\mu\nu}\,G_{(\xi\psi)}^{\rho\sigma}(p)\mathcal{P}^{\fth}_{22;\rho\sigma,\tau\delta}(p) &=& 
2\mathcal{P}^{\fth}_{22;\mu\nu,\tau\delta}(p)\label{prop_proj52v1}.
\eea
This leads  to a contradiction: from eq.\,\refe{prop_proj52v1} follows that  
$G_{(\xi\psi)}^{\rho\sigma}(p)\mathcal{P}^{\fth}_{22;\rho\sigma,\tau\delta}(p)$
cannot be zero but multiplying the both sides of the same equation  by $\mt^{\mu\nu}$ and using the property 
$\mt^{\mu\nu}\mathcal{P}^{\fth}_{22;\mu\nu,\tau\delta}(p)=0$ one can draw an opposite conclusion.
Hence, the quantity defined in eq.\,\refe{prop_proj52} does not obey eqs.\,\refe{propag_eqn} and cannot be a 
spin-$\ffh$ propagator.

The solution of equation  \refe{propag_eqn} where the parameters  $a$ and  $b$ are kept to be free is 
tedious. Here we confine ourselves by looking for a solution with a specific 
choice of the parameters  $a=-1$, $b=-1$ whereas $c$ is kept to be arbitrary. The independence of
$G_{(\psi\psi)}^{\tau\lambda;\rho\sigma}(p)$ on the latter parameter signifies that the auxiliary 
field does not contribute to the physical observables. We discuss this issue in Section \ref{Coupling}. 
With this specific choice of the free parameters  the resulting equations are  

\bea
[(\slash p -m )(\mt_{\mu\tau}\mt_{\nu\lambda} + \mt_{\mu\lambda}\mt_{\nu\tau})
    -\,(\gamma_\mu p_\nu \mt_{\lambda\tau} + \gamma_\nu p_\mu \mt_{\lambda\tau} 
      +\gamma_\lambda p_\tau \mt_{\mu\nu} + \gamma_\tau p_\lambda \mt_{\mu\nu} )
       +\left( \slash p\, +m \,\right) \mt_{\mu\nu}\mt_{\tau\lambda}\nonumber\\
-(\gamma_\mu p_\lambda \mt_{\nu\tau} + \gamma_\nu p_\lambda \mt_{\mu\tau} 	
      + \gamma_\mu p_\tau \mt_{\nu\lambda} +\gamma_\nu p_\tau \mt_{\mu\lambda} 
+  \gamma_\lambda p_\mu \mt_{\nu\tau} +\gamma_\tau p_\mu \mt_{\nu\lambda} 
      +\gamma_\lambda p_\nu \mt_{\mu \tau}  +\gamma_\tau p_\nu \mt_{\mu\lambda})\nonumber\\ 
+ (\gamma_\mu \slash p\gamma_\lambda   \mt_{\nu\tau}
      +\gamma_\nu \slash p\gamma_\lambda   \mt_{\mu\tau}
      +\gamma_\mu \slash p\gamma_\tau \mt_{\nu\lambda} 
      +\gamma_\nu \slash p\gamma_\tau \mt_{\mu\lambda} )
+m\,( \gamma_\mu\gamma_\lambda   \mt_{\nu\tau}
      +  \gamma_\nu\gamma_\lambda   \mt_{\mu\tau}\nonumber\\
      +  \gamma_\mu\gamma_\tau \mt_{\nu\lambda  }
      +  \gamma_\nu\gamma_\tau \mt_{\mu\lambda  } )
]
G_{(\psi\psi)}^{\tau\lambda;\rho\sigma}(p) + m\, c\, \mt_{\mu\nu}\,G_{(\xi\psi)}^{\rho\sigma}(p) = 
\mt_\mu^\rho\, \mt_\nu^\sigma+\mt_\mu^\sigma\, \mt_\nu^\rho,  \nonumber\\
\,m\, \mt_{\tau\lambda}  G_{(\psi\xi)}^{\tau\lambda}(p) - \frac{6\,c^2}{5}\left(\slash p +3m\right)(p)\,
   G_{(\xi\xi)}(p)=1.
\label{propag_eqn2}
\eea

The obtained spin-$\ffh$ propagator $G_{\ffh}^{\mu\nu;\rho\sigma}(p)=G_{(\psi\psi)}^{\mu\nu;\rho\sigma}(p)$ 
can be written as a decomposition in terms of projection operators as follows
\cite{Berends:1979rv}
\bea
G^\ffh_{\mu\nu;\rho\sigma}(p) &=& \frac{1}{p^2-m^2}
\left( 
\left (\slash p +m\right) \mathcal{P}^\ffh_{\mu\nu;\rho\sigma}(p)
 - \frac{p^2-m^2}{m^2}\left(\mathcal{D}^\fth_{\mu\nu;\rho\sigma}(p)+ \mathcal{D}^\foh_{\mu\nu;\rho\sigma}(p) 
\right) \right),
\label{propag_eqn3}
\eea
where $\mathcal{D}^\fth_{\mu\nu;\rho\sigma}(p)$ and  $\mathcal{D}^\foh_{\mu\nu;\rho\sigma}(p)$ 
stand for the contributions from the spin-$\fth$ and -$\foh$ sector of the spinor-vector representation, 
(see Appendix \ref{app_aux}).
As expected the propagator for the spin-$\ffh$ itself does not depend on the parameter $c$  related to
the  spinor field $\xi$. This observation also holds for arbitrary values of $a$ and $b$ in eq.\,\refe{propag_eqn}
 which we have checked by explicit calculations. The obtained propagator has a pole associated  
with the  spin-$\ffh$ part and so called 'off-shell' non-pole contributions 
coming from the lower spin components $\mathcal{D}^\fth_{\mu\nu;\rho\sigma}(p)$ and  
$\mathcal{D}^\foh_{\mu\nu;\rho\sigma}(p)$.

\section{\label{Coupling}Coupling to  higher spin fermions}
In the  case of the spin-$\ffh$  field in the spinor-tensor representation 
we deal with a system $(\psi_{\mu\nu},\xi)$ which contains auxiliary 
degrees of freedom. One can raise a question whether the unphysical degrees of freedom could
be eliminated from physical observables. Here we consider a simple case of spin-$\ffh$ resonance contribution to  
$\pi N$ scattering  which is valid for applications in hadron physics.   
 The corresponding $\pi NN^*_\ffh$  coupling can be chosen as follows
\bea
\mathcal{L}_{I}=\frac{{\rm g}_{\pi N N^*}}{4m_\pi^2}\,\left(\bar \psi_N(x),0\right)
\Gamma_{\mu\nu;\rho\sigma} 
\left[ \hat P_{(\psi\psi)}
 \left(\begin{array}{c}
\psi^{\rho\sigma}\nonumber\\
\xi \\
\end{array}\right)\right] 
\, \partial^\mu \partial^\nu\pi(x) + {\rm h.c.},
\label{coupling_gen} 
\eea 
where the nucleon field is written  as $\left(\bar \psi_N(x),0\right)$ which implies the absence of 
auxiliary fields in the final state. The operator
\bea
\hat P_{(\psi\psi)}=
\left(\begin{array}{cc}
1 & 0\nonumber\\
0 & 0 \\
\end{array}\right)
\eea
projects out the spin-$\ffh$ field and ensures that there is no coupling to $\xi$. 
Hence, only the spin-$\ffh$ component of the propagator $G^\ffh_{\mu\nu;\rho\sigma}(p)$,  
eq.\,\refe{propag_eqn3}, contributes to  physical observables at any order of perturbation theory. 
In Section \ref{FreeVector} we have demonstrated that the inclusion of auxiliary degrees of freedom 
in the vector field does not affect the physical observables. To our knowledge this  statement  
is not generally proven  for  the $(\psi_{\mu\nu},\xi)$ system beyond the pertubation expansion.
The reason is that the equation of the motion for massive spin-$\ffh$ field in the spinor-tensor representation 
is defined only in the presence of an auxiliary field. This is unlike the case
of the vector field where auxiliary degrees of freedom can be removed by  proper field transformations. 
Note that these degrees of freedom  contribute due to $\psi_{\mu\nu}-\xi$ mixing. This mixing takes 
place only between the spin-$\foh$ sector of the spinor-tensor and the auxiliary spinor fields, 
as pointed out in the previous Section.
One may therefore hope  that the use of a coupling  which suppresses the spin-$\foh$ contributions would 
also prevent the appearance of the auxiliary degrees of freedom in the physical observables in the non perturbative
regime.

 A possibility to remove unwanted degrees of freedom in a special case of the spin-$\fth$ fields has been 
demonstrated in \cite{Pascalutsa:1998pw,Pascalutsa:2000kd,Pascalutsa:1999zz}. The idea is based
on the observation that the Lagrangian of the Rarita-Schwinger fields for a specific choice of the 
parameter $A=-1$ maintains  gauge-invariance in the massless limit. Therefore the use of a gauge-invariant
coupling suppresses the contribution from the lower spin sector.

Guided by the results obtained in the spin-$\fth$ Rarita-Schwinger theory one
could expect that the lower spin terms of the spin-$\ffh$ propagator eq.\,\refe{propag_eqn3}
do not contribute to the physical observables as long as  a gauge invariant coupling  is used.
This however is not generally true for the spin-$\ffh$ fields in the spinor-tensor representation.
Such a conclusion can be drawn from the fact that the Lagrangian eq.\,\refe{applagr_nontr} is not invariant
under the gauge transformations $\psi_{\mu\nu}\to\psi_{\mu\nu} + \partial_\mu\xi_\nu + \partial_\nu\xi_\mu$
in a massless limit at any choice of parameters $a$ and $b$.
We show this more explicitely by exploring a general structure  of the gauge-invariant vertex function on the example
of the $\pi N$ scattering amplitude in the leading order of the perturbation expansion. The amplitude 
can in general be written in the form 
\bea
\mathcal{M}\sim\bar u_N(p')\left [\,
 \Gamma^{\mu\nu;\rho\sigma}(q)\,G_{\rho\sigma;\alpha\beta}^{\ffh}(q)
 \,\Gamma^{\mathrm{\dag}\,\alpha\beta;\lambda\tau}(q) \,\right] u_N(p) \,k'_\mu \, k'_\nu \,
k_\lambda \,k_\tau,
\label{tran_matrix} 
\eea
where  $p\,( k )$ and $p'(k')$ are momenta of the initial and final nucleon(pion) correspondingly; $q$ 
stands for the momentum of the resonance and depends on  the channel ($s$- or $u$- ) of interest. The gauge-invariant
coupling to the spin-$\ffh$ field imposes the following constraint on the vertex function   
$q_\rho\, \Gamma^{\mu\nu;\rho\sigma}(q) =q_\sigma \,\Gamma^{\mu\nu;\rho\sigma}(q) =0$. 
For the transition matrix to be free from any contribution from the lower spin sector the expression in the square
brackets in eq.\,\refe{tran_matrix} should be proportional to the spin-$\ffh$ projection operator
\bea
\left[\,
 \Gamma^{\mu\nu;\rho\sigma}(q)\,G_{\rho\sigma;\alpha\beta}^{\ffh}(q)
 \,\Gamma^{\mathrm{\dag}\,\alpha\beta;\lambda\tau}(q) \,\right]\sim \mathcal{P}^\ffh_{\mu\nu;\rho\sigma}(q).
\eea
This gives an additional constraint
$q_\mu\, \Gamma^{\mu\nu;\rho\sigma}(q) =q_\nu \,\Gamma^{\mu\nu;\rho\sigma}(q) =0$. 
The vertex function can be decomposed in terms
of the spin projection operators. There are only three operators 
$\mathcal{P}^\ffh_{\mu\nu;\rho\sigma}(q)$, $\mathcal{P}^\fth_{22;\,\mu\nu;\rho\sigma}(q)$, and
 $\mathcal{P}^\foh_{22;\,\mu\nu;\rho\sigma}(q)$ which fulfill the properties 
$q^\mu\, \mathcal{P}^\ffh_{\mu\nu;\rho\sigma}(q)=0$, $q^\nu\,\mathcal{P}^\fth_{22;\,\mu\nu;\rho\sigma}(q)=0$, 	 etc.
Hence, the  decomposition can be written as follows
\bea
 \Gamma_{\mu\nu;\rho\sigma}(q) = \alpha_1(\slash q)\, \mathcal{P}^\ffh_{\mu\nu;\rho\sigma}(q)+
 \alpha_2(\slash q)\, \mathcal{P}^\fth_{22;\,\mu\nu;\rho\sigma}(q) +\alpha_3(\slash q)\, 
\mathcal{P}^\foh_{22;\,\mu\nu;\rho\sigma}(q),
\label{vertex_gauge}
\eea
where the coefficients of the decomposition $\alpha_1(\slash q)$, $\alpha_2(\slash q)$, and  $\alpha_3(\slash q)$ are
polynomials  of m and $\slash q$. 
 Note, that $\mathcal{P}^\ffh_{\mu\nu;\rho\sigma}(q)$, $\mathcal{P}^\fth_{22;\,\mu\nu;\rho\sigma}(q)$, 
and $\mathcal{P}^\foh_{22;\,\mu\nu;\rho\sigma}(q)$ commute with $\slash q$. The spin-$\ffh$ propagator 
can  also be decomposed 
in terms of the spin projection operators. Due to the orthogonality properties of the projection operators 
only those terms in  $G_{\rho\sigma;\alpha\beta}^{\ffh}(q)$
contribute to the matrix element  eq.\,\refe{tran_matrix} which contain  $\mathcal{P}^\ffh_{\mu\nu;\rho\sigma}(q)$, 
$\mathcal{P}^\fth_{22;\,\mu\nu;\rho\sigma}(q)$, and $\mathcal{P}^\foh_{22;\,\mu\nu;\rho\sigma}(q)$ operators.
If the parameters $a$  and $b$ in eq.\,(\ref{appprop_matrix},\,\ref{propag_eqn}) could be chosen in such 
a way that the propagator does not contain the $\mathcal{P}^\fth_{22;\,\mu\nu;\rho\sigma}(q)$ and 
$\mathcal{P}^\foh_{22;\,\mu\nu;\rho\sigma}(q)$ operators the lower spin contributions to the matrix element would be
 suppressed. Such a situation is realized in the spin-$\fth$ Rarita-Schwinger theory for the special choice of the
free parameter $A=-1$, see \cite{Shklyar:2008kt} for discussion. For the spin-$\ffh$ fields eq.\,\refe{appspin52_eq2}
the contribution of the $\mathcal{P}^\fth_{22;\,\mu\nu;\rho\sigma}(q)$ projector can be suppressed by choosing $a=-1$.
As we already mentioned in  Section \ref{Free52} this parameter is associated with both spin-$\fth$ and -$\foh$
degrees of freedom whereas $b$ regulates only the spin-$\foh$ ones. Indeed the expression  
eq.\,(\ref{app_prop}) derived for $a=-1$, $b=-1$ does not have the 
$\mathcal{P}^\fth_{22;\,\mu\nu;\rho\sigma}(q)$ projector. One can ask whether  
$\mathcal{P}^\foh_{22;\,\mu\nu;\rho\sigma}(q)$ can also be removed from the free propagator.
The general conclusion is that the term $G_{(\psi\psi)}^{\tau\lambda;\rho\sigma}(p)$ 
being a solution of eq.\refe{propag_eqn} always has contributions from
$\mathcal{P}^\foh_{22;\,\mu\nu;\rho\sigma}(q)$. We have checked  this by explicite calculation for arbitrary values 
of parameter b. This conclusion is ultimately linked to the fact 
that the Lagrangian of the free spin-$\ffh$ fields \refe{applagr_nontr}  does not maintain gauge-invariance in the 
massless limit. The same conclusion has been also drawn in \cite{Berends:1979rv}. 
Therefore one can never remove the corresponding degrees of freedom from the transition matrix eq.\,\refe{tran_matrix} 
provided the vertex function is written in the form of eq.\,\refe{vertex_gauge}. 
 
The solution to the problem has been suggested in \cite{Shklyar:2008kt} where authors suggested
to utilize the  Rarita-Schwinger condition $\gamma_\mu \psi^{\mu\nu}=\gamma_\nu\psi^{\mu\nu}=0$ 
to constrain the interaction vertex. As a result the interaction vertex fulfills the condition 
$\gamma\cdot \Gamma=\Gamma\cdot\gamma=0$.  Applying this constraint to the decomposition 
eq.\,\refe{vertex_gauge} one can see that only the $\mathcal{P}^\ffh_{\mu\nu;\rho\sigma}(q)$ 
projector obeys the desired property and the decomposition reduces to 
 \bea
 \Gamma_{\mu\nu;\rho\sigma}(q) = \alpha_1(\slash q)\, \mathcal{P}^\ffh_{\mu\nu;\rho\sigma}(q).
 \label{vertex_gauge1}
\eea
Since the vertex function should be free from any singularities the minimal  power of $\slash q$ in the
function $\alpha_1(\slash q)$  should be of the fourth order. Then, the most simple coupling  
can be written as follows
\bea
\mcl^{\ffh}_{\pi N N^*} = \frac{{\rm g}_{\pi N N^*}}{m_\pi^2 m_R^4}\,\bar \psi_N(x)
\left[\Box^2\mathcal{P}^\ffh_{\mu\nu;\rho\sigma}(\partial) \psi^{\mu\nu}_{N^*}\right] \partial^\mu \partial^\nu
 \pi(x) + {\rm h.c.}
\label{coupling1}
\eea 
The use of $\mathcal{P}^\ffh_{\mu\nu;\rho\sigma}(\partial)$
ensures that only  the  spin-$\ffh$ part of the 
propagator contributes and the d'Alembert-operator squared
guarantees that no other singularities except the mass pole term $(p^2-m^2)^{-1}$ appear in the amplitude.
As a results the physical observables no longer depend on the arbitrary parameters $a$ and $b$ of the 
free Lagrangian. The $\pi N$ scattering amplitude eq.\,\refe{tran_matrix} then reads 
\bea
\mathcal{M}=\left(\frac{{\rm g}_{\pi N N^*}}{m_\pi^2}\right)^2
\bar u_N(p')\left [\, \left(\frac{q^2}{m_R^2}\right)^4
\mathcal{P}^\ffh_{\mu\nu;\lambda\tau}(q) 
 \,\right] u_N(p) \, k'^\mu \, k'^\nu \,
k^\lambda \,k^\tau,
\label{tran_matrix2} 
\eea
The coupling in eqs.\,(\ref{coupling1}) can be generalized for the 
fields $\psi_{N^*}^{\{\rho...\sigma\}}$ of the arbitrary spin $J$ as
\bea
\mcl^{ J}_{\pi N N^*} = \frac{{\rm g}_{\pi N N^*}}{(m_\pi)^{\frac{1}{2}(J-1/2)}}\,\bar \psi_N(x)
\left[\left(\frac{\Box}{m_R^2}\right)^{(J-\frac{1}{2})}
\mathcal{P}^J_{\{\mu...\delta;\rho...\sigma\}}(\partial) 
\psi_{N^*}^{\{\rho...\sigma\}}\right] 
 \{\partial^\mu\}...\{\partial^\delta\} \,\pi(x) + {\rm h.c.},
\label{coupling2}
\eea 
where the number of indices assigned to $\psi_{N^*}^{\{\rho...\sigma\}}$ and 
$\mathcal{P}^J_{\{\mu...\nu;\rho...\sigma\}}(\partial)$ depends on the chosen representation.
The coupling constructed in eqs.\,(\ref{tran_matrix2},\,\ref{coupling2}) ensures that physical observables
do not depend on the free parameters of the theory.

\section{\label{Summary} Summary}
In summary, we have investigated the  general properties of the free spin-$\ffh$ fields in the spinor-tensor 
representation. 
The Lagrangian is written in terms of spin-$\ffh$ and auxiliary fields  $(\psi_{\mu\nu},\xi)$
and coincides with that  suggested in the literature for  a specific choice of free parameters.
We demonstrate  that the Lagrangian in general depends on three arbitrary parameters; 
two of them are associated with the lower spin-$\fth$ and  -$\foh$ sector of the theory 
whereas the third one is linked to the auxiliary field $\xi$.

We deduce a free propagator of the system  which is given by a  2\,x\,2 matrix in the 
$(\psi_{\mu\nu},\xi)$ space. The diagonal elements stand for the propagation of the spin-$\ffh$ 
and $\xi$ fields whereas the non-diagonal ones correspond to  $\psi_{\mu\nu}-\xi$ mixing. 
The mixing takes place between the spin-$\foh$ sector of the spinor-tensor representation and 
an auxiliary spinor field.
  While the free propagator includes auxiliary degrees of freedom  
they do not contribute to the physical observables calculated within the perturbation theory 
provided there is no coupling to $\xi$.
 
  As an application to  hadron physics calculations, the interaction involving  
$(\psi_{\mu\nu},\xi)$ is discussed for the example of the $\pi N N^*_\ffh$ coupling.
The pure  spin-$\ffh$ propagator contains non-pole terms  which contribute
in the whole kinematical region. As we demonstrate invariance under gauge transformations
is not enough to remove these contributions. This is ultimately related  to  the fact that the free
Lagrangian of the   $(\psi_{\mu\nu},\xi)$ system does not maintain gauge invariance in the
massless limit for any choice of the free parameters.  
 The desired result can, however, be  obtained by constructing 
a coupling with higher order derivatives. In the latter case the amplitude of the $\pi N$ scattering 
does not depend on the arbitrary  parameters of the free Lagrangian. The suggested coupling is generalized to 
the Rarita-Schwinger fields  of any half-integer spin.

\begin{acknowledgments}
The work has been supported DFG, contract Le 435/7-1 and by SFB/TR16
Bonn-Giessen-Bochum {\em Subnuclear Structure of Matter}.
\end{acknowledgments}

\begin{appendix}
\section{\label{projectors}Spin projection operators for the spinor-tensor representation} 
The spin projection operators are taken from \cite{Berends:1979rv}. In the momentum space they are given 
by
\bea
\mathcal{P}^\ffh_{\mu\nu;\rho\sigma}(q)&=& \frac{1}{2}\left(\mathcal{P}^1_{\mu\rho}\,\,\mathcal{P}^1_{\nu\sigma}
+ \mathcal{P}^1_{\mu\sigma}\,\mathcal{P}^1_{\nu\rho}\right)
-\frac{1}{5}\,\mathcal{P}^1_{\mu\nu}\,\,\mathcal{P}^1_{\rho\sigma}\nonumber\\
&-&\frac{1}{10}\left(
                   {\ll \mathcal{P}}^1_\mu\,\,{\ll \mathcal{P}}^1_\rho\,\,\mathcal{P}^1_{\nu\sigma}
                +  {\ll \mathcal{P}}^1_\nu\,\,{\ll \mathcal{P}}^1_\rho\,\,\mathcal{P}^1_{\mu\sigma}
                +  {\ll \mathcal{P}}^1_\mu\,\,{\ll \mathcal{P}}^1_\sigma\,\,\mathcal{P}^1_{\nu\rho}
                +  {\ll \mathcal{P}}^1_\nu\,\,{\ll \mathcal{P}}^1_\sigma\,\,\mathcal{P}^1_{\mu\rho}
              \right),\nonumber\\
\nonumber\\
%
%
\mathcal{P}^\fth_{11;\mu\nu;\rho\sigma}(q)&=& 
 \frac{1}{2}\left(
                  \mathcal{P}^1_{\mu\rho}\,\,\mathcal{P}^0_{\nu\sigma}
                + \mathcal{P}^1_{\nu\rho}\,\,\mathcal{P}^0_{\mu\sigma}
                + \mathcal{P}^1_{\mu\sigma}\,\,\mathcal{P}^0_{\nu\rho}
                + \mathcal{P}^1_{\nu\sigma}\,\,\mathcal{P}^0_{\mu\rho}
             \right)
-\frac{1}{6\,q^2}\,\mathcal{O}_{\mu\nu}\,\,\mathcal{O}_{\rho\sigma},
\nonumber\\
\nonumber\\
%
%
\mathcal{P}^\fth_{22;\mu\nu;\rho\sigma}(q)&=& 
 \frac{1}{10}\left(
                   {\ll \mathcal{P}}^1_\mu\,\,{\ll \mathcal{P}}^1_\rho\,\,\mathcal{P}^1_{\nu\sigma}
                +  {\ll \mathcal{P}}^1_\nu\,\,{\ll \mathcal{P}}^1_\rho\,\,\mathcal{P}^1_{\mu\sigma}
                +  {\ll \mathcal{P}}^1_\mu\,\,{\ll \mathcal{P}}^1_\sigma\,\,\mathcal{P}^1_{\nu\rho}
                +  {\ll \mathcal{P}}^1_\nu\,\,{\ll \mathcal{P}}^1_\sigma\,\,\mathcal{P}^1_{\mu\rho}
              \right)
-\frac{2}{15}\,\mathcal{P}^1_{\mu\nu}\,\,\mathcal{P}^1_{\rho\sigma},
\nonumber\\
\nonumber\\
\mathcal{P}^\fth_{21;\mu\nu;\rho\sigma}(q)&=& 
-\mathcal{P}^\fth_{12;\rho\sigma;\mu\nu}(q)=
 \frac{1}{2\sqrt{5}\,q^2}\,
             \left(
                   q_\rho\,{\ll \mathcal{P}}^1_\mu\,\, \mathcal{P}^1_{\nu\sigma}
                +  q_\rho\,{\ll \mathcal{P}}^1_\nu\,\, \mathcal{P}^1_{\mu\sigma}
                +  q_\sigma \,{\ll \mathcal{P}}^1_\mu\,\, \mathcal{P}^1_{\nu\rho}
                +  q_\sigma \,{\ll \mathcal{P}}^1_\nu\,\, \mathcal{P}^1_{\mu\rho}
             \right)\,\slash q\nonumber\\
        &-&\frac{1}{3\,\sqrt{5}\,q^2}\, \mathcal{P}^1_{\mu\nu}\,\,\mathcal{O}_{\rho\sigma}\,\slash q,
\nonumber\\
\nonumber\\
\mathcal{P}^\foh_{11;\mu\nu;\rho\sigma}(q)&=& \mathcal{P}^0_{\mu\nu}\,\,\mathcal{P}^0_{\rho\sigma},
\nonumber\\
\nonumber\\
\mathcal{P}^\foh_{22;\mu\nu;\rho\sigma}(q)&=& \frac{1}{3}\, \mathcal{P}^1_{\mu\nu}\,\,\mathcal{P}^1_{\rho\sigma},
\nonumber\\
\nonumber\\
\mathcal{P}^\foh_{33;\mu\nu;\rho\sigma}(q)&=& \frac{1}{6\,q^2}\, \mathcal{O}_{\mu\nu}\,\,\mathcal{O}_{\rho\sigma},
\nonumber\\
\nonumber\\
\mathcal{P}^\foh_{21;\mu\nu;\rho\sigma}(q)&=& 
\mathcal{P}^\foh_{12;\rho\sigma;\mu\nu}(q)
=\frac{1}{\sqrt{3}}\, \mathcal{P}^1_{\mu\nu}\,\,\mathcal{P}^0_{\rho\sigma} ,
\nonumber\\
\nonumber\\
\mathcal{P}^\foh_{31;\mu\nu;\rho\sigma}(q)&=& 
-\mathcal{P}^\foh_{13;\rho\sigma;\mu\nu}(q)= 
\frac{1}{\sqrt{6}\,q^2}\, \mathcal{O}_{\mu\nu}
\,\,\mathcal{P}^0_{\rho\sigma}\,\slash q,
\nonumber\\
\nonumber\\
\mathcal{P}^\foh_{23;\mu\nu;\rho\sigma}(q)&=& 
-\mathcal{P}^\foh_{32;\rho\sigma;\mu\nu}(q)= 
\frac{-1}{3\sqrt{2}\,q^2}\, \mathcal{O}_{\rho\sigma}
\,\,\mathcal{P}^1_{\mu\nu}\,\slash q,
\label{app_proj}
\eea
where operators   $\mathcal{P}^1_{\mu\nu}$, $\mathcal{P}^0_{\mu\nu}$, ${\ll \mathcal{P}}^1_\mu$,
and $\mathcal{O}_{\mu\nu}$ are defined as

\bea
 \begin{array}{ll}
    \mathcal{P}^1_{\mu\nu}={\rm g}_{\mu\nu} - \frac{q_\mu q_\nu}{q^2}, \qquad 
   & {\ll \mathcal{P}}^1_\mu = \mathcal{P}^1_{\mu\nu}\gamma^\nu,\nonumber
\\ \nonumber\\
     \mathcal{P}^0_{\mu\nu}= \frac{q_\mu q_\nu}{q^2}, 
   & \mathcal{O}_{\mu\nu}= {\ll \mathcal{P}}^1_\mu q_\nu +q_\mu{\ll \mathcal{P}}^1_\nu.
 \end{array}
\eea
The projection operators eq.\,\refe{app_proj} fulfill the following properties: they satisfy orthogonality
conditions 
\bea
 \mathcal{P}^{J\!\!\!\!\!\!\qquad;\tau\lambda}_{ii;\mu\nu}(q)\,\, \mathcal{P}^{J'}_{ii;\tau\lambda;\rho\sigma}(q)
= \delta_{JJ'}\mathcal{P}^J_{ii;\mu\nu;\rho\sigma}(q)
\label{app_property}
\eea
and the sum rules
\bea
\mathcal{P}^\ffh_{\mu\nu;\rho\sigma}(q)
+\sum_{i=1}^{2}
 \mathcal{P}^\fth_{ii;\mu\nu;\rho\sigma}(q)
+\sum_{i=1}^{3}
 \mathcal{P}^\foh_{ii;\mu\nu;\rho\sigma}(q)
=\frac{1}{2}(\mt_{\mu\rho}\mt_{\nu\sigma}+\mt_{\mu\sigma}\mt_{\nu\rho}).
\label{app_property2}
\eea

\section{\label{app_lagr}Lagrangian for the free spin-$\ffh$ field} 
The functions $F_1(a)$, $F_2(a)$, $G_1(a,b)$, $G_2(a,b)$, and $B(a,b,c)$ of the free real parameters 
$a$, $b$, and $c$ used in the definition of the Lagrangian eq.\,\refe{applagr_nontr} read
\bea
F_1(a)&=&\frac{1}{4}(5a^2 +2a+1),\nonumber\\
F_2(a)&=&\frac{1}{8}(15a^2+10a+3),\nonumber\\
G_1(a,b)&=&\frac{5a^4-12a^3-20a^2-8a-4b^2-4\,b\,(7a^2+6\,a+1)-1}{2(3a+1)^2},\nonumber\\
G_2(a,b)&=&\frac{-15a^4+18a^2+8a+12b^2+6\,b\,(5a^2+6\,a+1)+1}{2(3a+1)^2},\nonumber\\
B(a,b,c)&=&-\frac{24\,c^2\,(3a+1)^2}{5(5\,a^2+6\,a+4\,b+1)^2}.
\label{applagr_param}
\eea
Using the variational principle one obtains two equations of motion 
eqs.\,(\ref{appspin52_eq1},\,\ref{appspin52_eq2}). 
Here we show that all Rarita-Schwinger constraints \cite{Rarita:1941} can be obtained from these equations.
By multipling the  eq.\,\refe{appspin52_eq1} by $\mt^{\mu\nu}$, $\gamma^\mu p^\nu$ and $p^\mu p^\nu$ and 
making summation we get 
\bea
\big( \mt_{\rho\sigma}(\slash\, p (b-2F_1(a)+2G_1(a,b)+1) + m(\,2F_2(a)+2G_2(a,b)-1))\nonumber\\
+2(\gamma_\sigma p_\rho + \gamma_\rho p_\sigma)(a+b+ F_1(a))\big)\psi^{\rho\sigma} + 2\,c\, m\xi =0,
\label{app_eq1}\\
\nonumber\\
\big((2 a + 5\,b +2F_1(a)+G_1(a,b))\, p^2 \mt_{\rho\sigma}
+ m\, \slash p \,(G_2(a,b)-2F_2(a)) \mt_{\rho\sigma}\nonumber\\
+(-m+ 6\,m F_2(a) + (b - 1 + 4F_1(a)\,)\,\slash p \,
  )(\gamma_\rho p_\sigma + \gamma_\sigma p_\rho)\nonumber\\
+ (12a + 4 )\,p^\rho p^\sigma\big)\psi^{\rho\sigma} + c\,m \slash p \,\xi=0,\label{app_eq2}\\
\nonumber\\
\big(2((2a+1)\,\slash p-m)p^\rho p^\sigma
+(p^2(
2a + b +2F_1(a)) +2m F_2(a)\, \slash p\, 
)(\gamma^\rho p^\sigma +\gamma^\sigma p^\rho) \nonumber\\
+p^2((2b+G_1(a,b))\slash p + mG_2(a,b))\mt^{\rho\sigma}\big)\psi_{\rho\sigma}
+c\,m\,p^2\,\xi=0,
\label{app_eq3}
\eea
correspondingly. Expessing $(\gamma_\sigma p_\rho + \gamma_\rho p_\sigma)\psi^{\rho\sigma}$ 
from the first equation \refe{app_eq1} and substituing it in to eqs. (\ref{app_eq2},\,\ref{app_eq3}) 
we obtain

\bea
\Big(4 \,[3  a+1] [a+b+F_1(a)] \,p^{\rho }p^{\sigma } 
+\frac{1}{2}\big(-[6  F_2(a)-1] [2  F_2(a)+2  G_2(a,b)-1]m^2+
    2 \slash p\,  [b+G_1(a,b)-2 F_2(a) \nonumber \\
(a+3b+3G_1(a,b)+1)+F_1(a)(1-3G_2(a,b))
+ (a+1) G_2(a,b)]  m+p^2  [4\, a^2+14\,a\,b+9\,b^2+12F_1^2(a)\nonumber\\
+ 2F_1(a)  (4  a+6  b-3  G_1(a,b)-3)+2\,(a+1) \, G_1(a,b)+1]\big)\mt^{\rho \sigma }   
\Big)\psi_{\rho\sigma}\nonumber\\
 + c\, m\, \big([1-6  F_2(a)] m+\slash p \, [a-3 F_1(a)+1]\big)\xi=0
\label{app_eq4}
\eea


\bea
\Big(2[a+b+F_1(a)][(2 a+1) \slash p -m]  p^{\rho }  p^{\sigma }  
+\frac{1}{2}  
\big(-2 F_2(a)  [\,2  F_2(a)+2G_2(a,b)-1] m^2\slash p \nonumber\\
     -p^2  (\slash p  (-3  b^2-2  a  b+b-4  F_1^2(a)+2a+2a\, G_1(a,b)
         +  F_1(a)  (-4  a-4b+2  G_1(a,b )+2))\nonumber\\
+ m  (2G_2(a,b)  a-2 a-b+F_2(a)  (4a + 4b+4G_1(a,b)+2)+
           2F_1(a)(G_2(a,b)-1)))\big)\mt^{\rho \sigma } \Big)\psi_{\rho\sigma} \nonumber\\
+c\,  m\, \big(-[a+F_1(a)]\,p^2  -2  m  F_2(a)\,\slash p \,\big)\xi=0.
\label{app_eq5}
\eea
Now multiplying eq.\refe{app_eq4} from left by  $(6a+2)^{-1}((2a+1)\slash p -m)$, substracting 
it from eq.\refe{app_eq5} and using the definitions  \refe{applagr_param} we have
\bea
-\frac{c\,m(3a+1)}{8}  \left(    (5m^2+3p^2)\,\xi 
      + \frac{3\,c\,m}{B(a,b,c)}
   (\slash p -3m)\  \mt^{\rho \sigma } \psi_{\rho\sigma}
         \right)=0.
\label{app_eq6}
\eea
From eq.\,(\ref{appspin52_eq2}) eq.\,(\ref{app_eq6}) we obtain 
\bea
(3a+1)\,c\, m^3 \xi=0.
\label{app_eq6a}
\eea
which means that the auxiliary field is vanishes provided $a\neq-\frac{1}{3}$. Having $\xi= 0$
the remaing constraints 	
\bea
(\gamma_\mu p_\nu + \gamma_\nu p_\mu)\,\psi^{\mu\nu}&=&0,\nonumber\\
p_\mu p_\nu \,\psi^{\mu\nu}&=&0,\nonumber\\
\mt_{\mu\nu}\,\psi^{\mu\nu}&=&0,
\label{app_eq7}
\eea  
can be easily derived from eqs.(\ref{app_eq1}-\ref{app_eq3}) and eq.\refe{appspin52_eq2}. 

Now multiplying 
eq.\,(\ref{appspin52_eq1}) from left by $\gamma^\nu$ and $p^\nu$ and using  eqs.\,\refe{app_eq7}
we have two equations  

\bea
 ((a+6F_1(a)-1)\,\slash p\,+ (6  F_2(a)-1)m )(\gamma^\sigma  \mt^{\mu \rho }+\gamma ^\rho   
\mt^{\mu \sigma })\psi_{\rho\sigma} 
 + 2(3a+1)(  \mt^{\mu \rho }  p^{\sigma }+ p^{\rho }  \mt^{\mu \sigma })\psi_{\rho\sigma} =0,
\label{app_eq8}
\eea
\bea
((a+1) \slash p-m)( p^{\sigma }\mt^{\mu \rho } + p^{\rho }  \mt^{\mu \sigma }  )\psi_{\rho\sigma}
+(p^2  (a+F_1(a))+ m \slash p F_2(a)) )(\gamma^\sigma \mt^{\mu \rho }+ \gamma^\rho \mt^{\mu \sigma }  )
\psi_{\rho\sigma}=0.  
\label{app_eq9}
\eea
Again, multiplying eq.\,\refe{app_eq8} by $(6a+2)^{-1}((a+1)\slash p -m)$, substracting the resulting equation
from eq.\,\refe{app_eq9} and using definitions \refe{applagr_param} we get 
\bea
(\gamma^\sigma \mt^{\mu\rho}+ \gamma^\rho \mt^{\mu\sigma} )\,\psi_{\rho\sigma}=0,
\label{app_eq10}
\eea
provided $a\neq-\frac{1}{3},\,-\frac{1}{2}$. Then the constraint  
\bea
(p^\sigma \mt^{\mu\rho}+ \gamma^\rho \mt^{\mu\sigma} )\,\psi_{\rho\sigma}=0
\label{app_eq11}
\eea
immediately follows from eqs.\,(\ref{app_eq8},\,\ref{app_eq10}). Having  $\gamma^\rho\psi_{\rho\sigma}=0$, 
$p^\rho\psi_{\rho\sigma}=0$,  $\xi=0$, and $\mt^{\rho\sigma}\psi_{\rho\sigma}=0$ the equation \refe{appspin52_eq1} reduces
to the Dirac equation $(\slash p -m)\psi_{\rho\sigma}=0$. Finally we have shown that all Rarita-Schwinger constraint
can be obtained from eqs.\,(\ref{appspin52_eq1},\,\ref{appspin52_eq2}). Hence, the function $\psi_{\rho\sigma}$ obeying 
these equations describes the field with spin-$\ffh$.

\section{\label{app_aux} Spin-$\ffh$ propagator}  
The solution of the equation \ref{propag_eqn2} can be written in form
\bea
G^\ffh_{\mu\nu;\rho\sigma}(p) &=& \frac{1}{p^2-m^2}
\left( 
\left (\slash p +m\right) \mathcal{P}^\ffh_{\mu\nu;\rho\sigma}(p)
 + \frac{p^2-m^2}{m^2}\left(\mathcal{D}^\fth_{\mu\nu;\rho\sigma}(p)+ \mathcal{D}^\foh_{\mu\nu;\rho\sigma}(p) 
\right) \right),\nonumber\\
G_{(\psi\xi)}^{\rho\sigma}(p) &=& \frac{1}{64\,m^3c}\left(
(\slash p+m)(2(\gamma^\rho p^\sigma + \gamma^\sigma p^\rho) +5m \mt^{\rho\sigma})
+6(p^2-m\slash p)\mt^{\rho\sigma} 
-16p^\rho p^\sigma\right),
\label{app_prop}
\eea
where the lower spin-$\fth$, -$\foh$ parts $\mathcal{D}^\fth_{\mu\nu;\rho\sigma}(p)$ and $\mathcal{D}^\foh_{\mu\nu;\rho\sigma}(p)$ are
\bea
\mathcal{D}^\fth_{\mu\nu;\rho\sigma}(p)=
 -\frac{4}{5}(\slash p +m)\mathcal{P}^\fth_{11;\,\mu\nu;\rho\sigma}(p) 
 + \frac{m}{\sqrt{5}}
     \left (
      \mathcal{P}^\fth_{12;\,\mu\nu;\rho\sigma}(p) +\mathcal{P}^\fth_{21;\,\mu\nu;\rho\sigma}(p) 
     \right )
\label{app_d32}\\
\mathcal{D}^\foh_{\mu\nu;\rho\sigma}(p) = \frac{1}{80\, m^2}
 \biggl(
  -\frac{3}{8} \biggl[\left( (73 m^2-12 p^2)\slash p + 3 m(27m^2-8p^2 )\right)\mathcal{P}^\foh_{11;\,\mu\nu;\rho\sigma}(p) \nonumber\\
                    + \left( (35m^2 - 36p^2)\slash p - m(13m^2 + 96p^2)\right )\mathcal{P}^\foh_{22;\,\mu\nu;\rho\sigma}(p) \nonumber\\
                  -\sqrt{3} \left( (43m^2 -12 p^2)\slash p + m(47m^2 - 28p^2)\right )(\mathcal{P}^\foh_{12;\,\mu\nu;\rho\sigma}(p)
                                                                                 + \mathcal{P}^\foh_{21;\,\mu\nu;\rho\sigma}(p))\biggl]
                                                                                                                               \nonumber\\ 
               - \left( (16m^2 + 3p^2)\slash p + m(16m^2 - 15p^2)\right )\mathcal{P}^\foh_{33;\,\mu\nu;\rho\sigma}(p)\nonumber\\
              + \frac{9}{2\sqrt{6}}(3m^2 -2p^2)\,\slash p\,
                          (  \mathcal{P}^\foh_{13;\,\mu\nu;\rho\sigma}(p) - \mathcal{P}^\foh_{31;\,\mu\nu;\rho\sigma}(p))
                                                                                           \nonumber\\ 
               + \frac{3m}{2\sqrt{6}} (64m^2 -21p^2)
                          (  \mathcal{P}^\foh_{13;\,\mu\nu;\rho\sigma}(p) + \mathcal{P}^\foh_{31;\,\mu\nu;\rho\sigma}(p))
                                                                                           \nonumber\\ 
              + \frac{9}{2\sqrt{2}}(3m^2 +2p^2)\,\slash p\,
                          (  \mathcal{P}^\foh_{23;\,\mu\nu;\rho\sigma}(p) - \mathcal{P}^\foh_{32;\,\mu\nu;\rho\sigma}(p))
                                                                                           \nonumber\\ 
              - \frac{m}{2\sqrt{2}}(64m^2 -69p^2)
                          (  \mathcal{P}^\foh_{23;\,\mu\nu;\rho\sigma}(p) +\mathcal{P}^\foh_{32;\,\mu\nu;\rho\sigma}(p))
                                                                                            \biggl).
\label{app_d12}
\eea

\end{appendix}

\bibliographystyle{h-physrev3}
\bibliography{tau}

\begin{thebibliography}{10}

\bibitem{Rarita:1941}
W.~Rarita and J.~S. Schwinger,
\newblock Phys. Rev. {\bf 60}, 61 (1941).

\bibitem{Fierz:1939ix}
M.~Fierz and W.~Pauli,
\newblock Proc. Roy. Soc. Lond. {\bf A173}, 211 (1939).

\bibitem{Pascalutsa:1998pw}
V.~Pascalutsa,
\newblock Phys. Rev. {\bf D58}, 096002 (1998), hep-ph/9802288.

\bibitem{Shklyar:2008kt}
V.~Shklyar and H.~Lenske,
\newblock Phys. Rev. {\bf C80}, 058201 (2009), 0812.4435.

\bibitem{Krebs:2008zb}
H.~Krebs, E.~Epelbaum, and U.-G. Meissner,
\newblock (2008), 0812.0132.

\bibitem{Schwinger:1973rv}
J.~S. Schwinger,
\newblock PARTICLES, SOURCES AND FIELDS, Reading 1973, 459p.

\bibitem{Berends:1979rv}
F.~A. Berends, J.~W. van Holten, P.~van Nieuwenhuizen, and B.~de~Wit,
\newblock Nucl. Phys. {\bf B154}, 261 (1979),
\newblock Note, that there is an obvious missprint in definition of the spin
  $\mathcal{P}^\fth_{22\rho\sigma;\tau\delta}(p)$ projector in (A.1): the last
  term $(-1/15)\theta_{\mu\nu}\theta_{\rho\sigma}$ schould read
  $(-2/15)\theta_{\mu\nu}\theta_{\rho\sigma}$.

\bibitem{David:1995pi}
J.~C. David, C.~Fayard, G.~H. Lamot, and B.~Saghai,
\newblock Phys. Rev. {\bf C53}, 2613 (1996).

\bibitem{Renard:1972vv}
Y.~Renard,
\newblock Nucl. Phys. {\bf B40}, 499 (1972).

\bibitem{Zetenyi:2002jy}
M.~Zetenyi and G.~Wolf,
\newblock Heavy Ion Phys. {\bf 17}, 27 (2003), nucl-th/0202047.

\bibitem{shklyar:2004a}
V.~Shklyar, G.~Penner, and U.~Mosel,
\newblock Eur. Phys. J. {\bf A21}, 445 (2004), nucl-th/0403064.

\bibitem{Weinberg:1980kq}
S.~Weinberg and E.~Witten,
\newblock Phys. Lett. {\bf B96}, 59 (1980).

\bibitem{Pascalutsa:1999zz}
V.~Pascalutsa and R.~Timmermans,
\newblock Phys. Rev. {\bf C60}, 042201 (1999), nucl-th/9905065.

\bibitem{Pascalutsa:2000kd}
V.~Pascalutsa,
\newblock Phys. Lett. {\bf B503}, 85 (2001), hep-ph/0008026.

\end{thebibliography}

\end{document}